\documentstyle[twocolumn,aps,prl,psfig,floats]{revtex}

\begin{document}

\draft

\twocolumn[\hsize\textwidth\columnwidth\hsize\csname @twocolumnfalse\endcsname

\title{Long-range antiferromagnetic order in the S=1 chain compound LiVGe$_2$O$_6$ }

\author{M.D. Lumsden$^1$, G.E. Granroth$^1$, D. Mandrus$^{1,2}$, 
S.E. Nagler$^1$,
J.R. Thompson$^{1,2}$, J.P. Castellan$^3$ and B.D. Gaulin$^3$}
\address{$^1$ Oak Ridge National Laboratory, P.O. Box 2008, Oak Ridge, 
TN, 37831 U.S.A.}
\address{$^2$ Department of Physics, University of Tennessee, Knoxville, TN,
37996-1200 U.S.A.}
\address{$^3$ Department of Physics and Astronomy, McMaster University,\\
Hamilton, Ontario, L8S 4M1, Canada}
\maketitle

\begin{abstract}
The phase transition in the compound LiVGe$_2$O$_6$ has been proposed as 
a unique example of a spin-Peierls transition in an S=1 antiferromagnetic
chain.  We report neutron and x-ray diffraction measurements of LiVGe$_2$O$_6$
above and below the phase transition at T=24 K.  No evidence is seen for any
structural distortion associated with the transition.  The neutron results
indicate that the low temperature state is antiferromagnetic, driven by
ferromagnetic interchain couplings.

\end{abstract}


]

\narrowtext

The physics of one-dimensional (1D) 
spin systems has been a field of considerable
theoretical and experimental interest in recent years.  The enhanced 
importance of quantum fluctuations in these systems manifests itself in
many exotic phenomena, two of which are the Haldane gap 
in integer spin chains and the spin-Peierls transition in 1D half-integer spin
systems.  Haldane predicted in 1983 \cite{Haldane}, 
based on field-theoretical models, that Heisenberg, antiferromagnetic (AF),
integer spin chains possess a spectrum of magnetic excitations in which
lowest lying excited states are separated by a finite energy gap 
from the ground state.  This remarkable result contrasts the
half-integer case where the excitation spectrum is gapless.  However, 
in the presence of magnetoelastic coupling, the energy of 1D, AF,
half-integer spin systems can be lowered by a dimerization of the uniform
chains.  The resulting alternating spin chains possess a gapped 
magnetic excitation spectrum separating a singlet ground state from the
triplet of excited states.  This structural phase transition, driven
by gains in magnetic energy, from the uniform to dimerized spin chain is known
as the spin-Peierls transition \cite{spinpeierls}.  

Experimentally, the Haldane gap has been observed in numerous S=1 chain systems
including CsNiCl$_3$ \cite{Buyers}, AgVP$_2$S$_6$ \cite{Mutka},
YBaNiO$_5$ \cite{Xu}, Ni(C$_2$H$_8$N$_2$)$_2$NO$_2$ClO$_4$ \cite{Renard}
and Ni(C$_3$H$_{10}$N$_2$)$_2$N$_3$ClO$_4$ \cite{Zheludev}.
The spin-Peierls transition has 
fewer experimental realizations most of which are organic, such as 
Cu(Au)BDT \cite{spinpeierls} and MEM(TCNQ)$_2$ \cite{spinpeierls}
with the only known inorganic example being CuGeO$_3$ \cite{Hase}.  
Recently, results were published on a new S=1, chain
material LiVGe$_2$O$_6$ \cite{Millet}
which were interpreted as being a remarkable union of 
these two, seemingly unrelated, phenomena.  Powder susceptibility
measurements showed a rapid decrease on passing through a transition 
temperature of about 22 K reminiscent of the isotropic decrease in 
magnetization, associated with the opening of a singlet-triplet gap in the
excitation spectrum, which accompanies a spin-Peierls 
transition \cite{spinpeierls}.  However, the
pre-existing gapped excitation spectrum normally precludes the 
possibility of a spin-Peierls phase transition in integer spin chain systems.
It was suggested \cite{Millet} that the Haldane gap
was closed in LiVGe$_2$O$_6$ due to a novel, anomalously large 
biquadratic exchange
interaction and this now gapless integer spin chain system undergoes 
a spin-Peierls phase transition at T=22 K \cite{Millet}.

However, very recent $^7$Li nuclear magnetic resonance (NMR) \cite{Gavilano}
experiments produced results
which seemed to contradict this interpretation.  These measurements indicated
that the observed phase transition was
not a spin-Peierls transition and that the low temperature
phase was antiferromagnetic in nature.  Their results are also interpreted
as being consistent with a discontinuous phase transition.
In addition, these measurements suggest unusual
behavior in the ordered phase including a gapped excitation spectrum with
a gap $\Delta$/$k_B$=83 K \cite{Gavilano}, roughly twice the 
observed coupling constant J/$k_B$$\approx$45 K \cite{Millet},
despite an anisotropy D/$k_B$ of less than 20 K \cite{Millet}.  
The authors of this \cite{Gavilano} and a recent theory
paper \cite{Mila} discuss possible mechanisms for the low-temperature
properties of LiVGe$_2$O$_6$ resulting from
splitting of the three $t_{2g}$ orbitals which accomodate the two V$^{3+}$
electrons. This splitting results from a small trigonal 
distortion \cite{Millet} and separates the $d_{xz}$ orbital from the
low-lying d$_{xy}$, d$_{yz}$ doublet.
Two scenarios are suggested, the first of which
leads to a reduced intrachain coupling and subsequent enhancement
of biquadratic and next near-neighbor interactions \cite{Gavilano,Mila}.
The second scenario leads to increased importance of orbital degrees of 
freedom \cite{Gavilano} and possible orbital ordering.  However, until
now, the precise nature of the ordered state was unclear.

To elucidate the nature of the low-temperature ordered state, 
we have performed temperature dependent neutron diffraction
studies, and complimentary x-ray diffraction studies, above and below 
the transition temperature.  These measurements indicate a commensurate,
antiferromagnetic, long-range ordered low-temperature phase and provide
no evidence for a structural distortion associated with the transition.

LiVGe$_2$O$_6$ crystallizes in a monoclinic unit cell with space group
P2$_1$/c and room temperature lattice constants a=9.863(4) \AA, b=8.763(2) \AA,
c=5.409 \AA, and $\beta$=108.21(1)$^\circ$ \cite{Millet}.  
The structure is composed of chains of edge sharing 
VO$_6$ octahedra which are connected by GeO$_4$ tetrahedra.  Previous 
measurements \cite{Millet,Gavilano} have indicated that the V-ions 
exist in a 3+ oxidation state with an effective S=1 magnetic moment.

\begin{figure}[bt]
\centerline{
\psfig{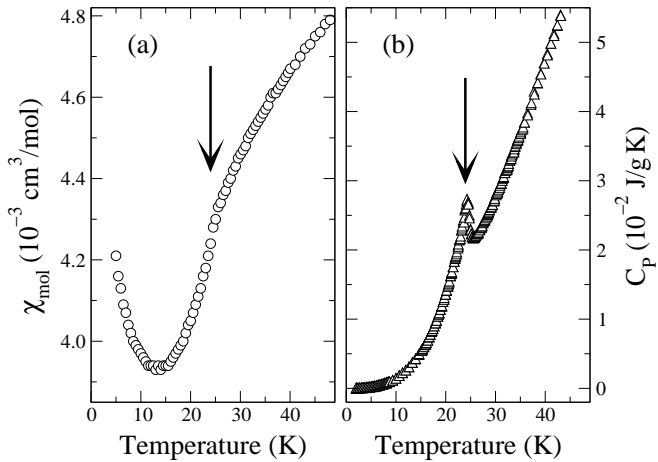}}
\vspace{2ex plus 1ex minus 0.5ex}
\caption{
DC molar susceptibility (a) and specific heat (b) as a function of 
temperature for temperatures in the vicinity of the transition temperature.
The DC magnetization was measured in an applied field of 2.5 kOe while the 
specific heat was a zero-field measurement.  Both measurements clearly 
show the presence of a phase transition at a temperature of about 24 K
indicated by the arrows.
}
\label{fig1}
\end{figure}

Polycrystalline samples of LiVGe$_2$O$_6$ were prepared 
using a slightly different
method than reported by Millet $et$ $al$.\cite{Millet}.  
Stoichiometric quantities
of GeO$_2$, Li$_2$O, V$_2$O$_5$, and V metal were sealed in evacuated 
Pt tubes and heated for 2 days at 900 $^\circ$C.  The samples were single
phase according to powder x-ray diffraction.
To reproduce the magnetization results reported previously, a small amount
of powder was loaded in a Quantum Design SQUID magnetometer and the DC
magnetization was measured as a function of temperature in an applied 
magnetic field of 2.5 kOe; the magnetization varied linearly with H
for fields up to 65 kOe .  The results are shown in Fig. 1(a) for temperatures
up to 50 K and clearly demonstrate the same rapid decrease in
magnetization, below a transition temperature of about 24 K, reported
previously \cite{Millet,Gavilano}.  The Curie tail below 10 K corresponds
to less than 40 ppm of 3d impurity content.  
In addition, specific heat measurements
were performed on a 13.7 mg cold-pressed pellet of LiVGe$_2$O$_6$ using a 
commercial heat-pulse, Quantum design calorimeter, the results of 
which are shown in Fig. 1(b).  These 
results show a clear phase transition at a temperature of about 24 K, 
consistent with the magnetization results.  

For the neutron diffraction measurements, approximately 10g of 
LiVGe$_2$O$_6$ powder was loaded in an aluminum sample
can in the presence of He exchange gas and attached to the cold finger
of a closed-cycle He refrigerator.  The neutron diffraction measurements 
were performed on the HB1 triple-axis spectrometer at the High Flux Isotope
Reactor, Oak Ridge National Laboratory.  Neutrons
of wavelength 2.44 \AA~ were selected using a pyrolitic graphite (PG) (002)
monochromator crystal and elastically scattered neutrons were measured 
using a second PG (002) analyzer crystal.
To prevent higher-order contamination, a graphite 
filter was placed in the incident beam.  The intensity of diffracted
neutrons was measured for scattering angles between 5$^\circ$ and 100$^\circ$ 
for temperatures of 50 K and 11 K, well above and below the transition 
temperature, respectively.  

The 50 K data was refined using Rietveld analysis and the observed pattern
was found to be consistent with the previously determined 
crystal structure \cite{Millet}
with low temperature lattice constants of a=9.800 \AA, b=8.709 \AA, 
c=5.364 \AA, and $\beta$=108.21$^\circ$.  

\begin{figure}[bt]
\centerline{
\psfig{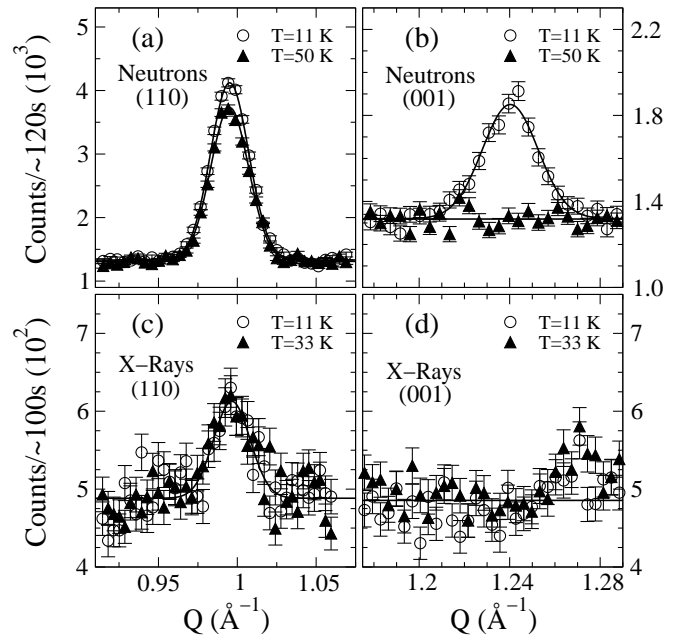}}
\vspace{2ex plus 1ex minus 0.5ex}
\caption{
The upper panels show powder neutron diffraction data for the (110) and (001)
reciprocal lattice positions at temperatures above and below T$_N$.
The lower panels show identical scans performed using x-ray 
diffraction techniques.  The neutron diffraction data indicates the 
presence of additional scattering below the transition temperature.  The
absence of excess scattering seen in the x-ray diffraction results confirms
the magnetic nature of the scattering. 
}
\label{fig2}
\end{figure}

A portion of the resulting diffraction pattern is shown in the two upper panels
of Fig. 2 for temperatures of 11 K and 50 K.  
We see clear, excess scattering at positions which can be 
described by Miller indices (110) and (001).  
These two reflections were the most pronounced but
other, weaker reflections were seen at (111) and (220) positions
with possible reflections at (-1-11) and (020).  No 
additional scattering was observed for reflections with scattering angles
in excess of about 50$^\circ$.  This enhacement of the scattering at low angles
is consistent with it being magnetic in origin.  

To further confirm the magnetic nature of
the scattering, the measurements were repeated using x-ray diffraction
techniques.  Approximately 2g of LiVGe$_2$O$_6$ powder was pressed 
into a pellet and loaded
in a Be can in the presence of He exchange gas.  This can was then connected
to the cold finger of a closed-cycle He refrigerator and mounted
in a Huber four-circle diffractometer.  The incident radiation was 
Cu-K$_\alpha$ x-rays from an 18 kW rotating anode x-ray generator which were
further monochromatized by Bragg reflection from a PG monochromator
crystal.  To allow comparison with the neutron diffraction results,
the intensity of diffracted x-rays was measured for scattering angles
between 5$^\circ$ and 70$^\circ$ at 
temperatures of 33 K and 11 K, again above and below the 
transition temperature.  The resulting data in the vicinity of the (110) and
(001) reflections is shown in the lower two panels of Fig. 2.  Clearly, 
no additional scattering is seen on passing through the transition temperature
and, in fact, no excess scattering was observed at any angle.  This further
confirms the magnetic nature of the excess scattering observed in the 
neutron diffraction experiment.  This coupled with the resolution limited 
nature of the excess neutron intensity, indicates that the low
temperature phase is, in fact, a long-range ordered, commensurate magnetic
state.

The temperature dependence of the (110) and (001) magnetic Bragg reflections
was measured from 11 K to about 40 K and the resulting data are shown in the
upper panel of Fig. 3.  For reference, we have plotted the temperature 
derivative of the molar susceptibility and replotted the specific heat data
in the lower panel of Fig. 3.  The excess neutron intensity clearly disappears
on passing through the transition temperature indicating its
association with the phase transiton.  The temperature 
dependence can be well described by a power law in the reduced temperature
suggestive of a continuous phase transition.  In addition, no hysteresis 
has been observed in either the magnetization or the specific heat and, thus,
in contrast to the NMR results \cite{Gavilano}, 
we conclude that the transition is continuous in nature.
While we can rule out the possibility of the transition being strongly
first-order in nature,  it is impossible to 
rule out a weakly first order phase transition from the existing data.

\begin{figure}[bt]
\centerline{
\psfig{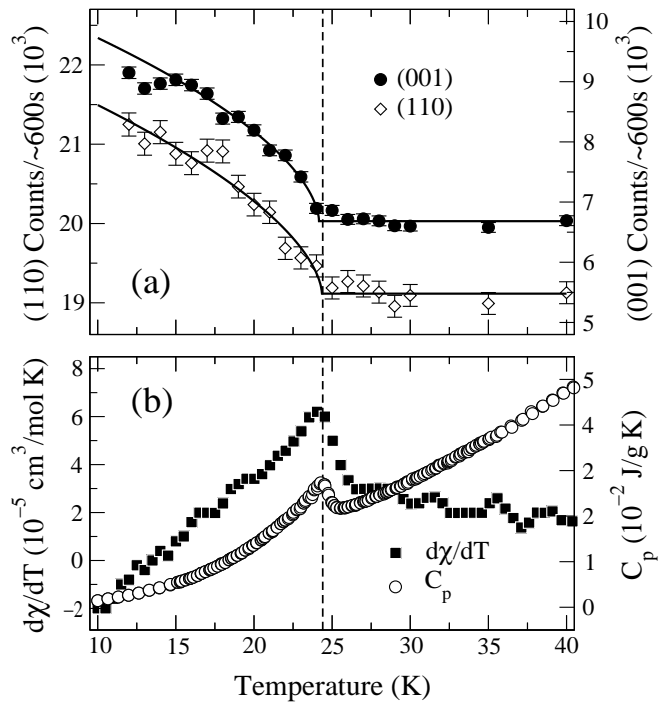}}
\vspace{2ex plus 1ex minus 0.5ex}
\caption{
The upper panel (a) shows the temperature dependence of the diffracted
neutron intensity for the (001) and (110)
magnetic Bragg reflections.  The solid lines represent fits to a power
law in the reduced temperature, $t$=$|T-T_N|/T_N$.
 For reference, the lower panel (b) shows
the temperature derivative of the susceptibility and the specific heat.
The dashed line represents the transition temperature (about 24 K).  This 
clearly shows that the extra magnetic neutron scattering disappears upon
passing through the transition temperature.
}
\label{fig3}
\end{figure}

Despite the small number of observed reflections, we have sufficient 
information to determine 
the spin arrangement in the ordered state.  As all reflections can be 
indexed with integer indices, the magnetic unit cell is 
identical in size to the nuclear unit cell and, hence, contains four
magnetic V$^{3+}$ sites. The observed magnetic peak intensities, coupled with
the absence of (100) and (010) reflections, are most consistent with the
arrangment of spins shown in Fig. 4 where we have superimposed the spins
on the known crystal structure.  In comparing various magnetic structures, 
an analytical approximation for the V$^{3+}$ magnetic form factor was employed
\cite{Brown}.
The observed spin structure has spins aligned
antiferromagnetically along the chain direction with ferromagnetic alignment
between neighboring chains suggestive of ferromagnetic interchain coupling.
Unfortunately, the limited amount of data makes it impossible to uniquely
determine the spin direction but the strength of the (001) reflection indicates
a small spin component along that direction.  By 
considering several plausible spin directions an 
ordered moment ($\mu$=g$\mu_B$$<$S$>$) of 1.14(8) $\mu_B$ was extracted.  
This is reduced
from the expected value of 1.79 $\mu_B$ (for g=1.79 \cite{Millet}
and $<$S$>$=1) and this difference is likely due to the importance of quantum
fluctuations in the low-dimensional system.  In fact, the observed ordered
moment is similar to that seen in other Haldane chain systems which
order magnetically: CsNiCl$_3$ is found to possess an ordered moment of
1.05 $\mu_B$ \cite{Minkiewicz}
and the set of systems RBaNiO$_5$, which if R $\not=$ Y order magnetically,
display an ordered moment of about 1.1 $\mu_B$ \cite{Zheludev2}.

The observed ordering in LiVGe$_2$O$_6$ is very simple and, as mentioned
above, exhibits properties consistent with other Haldane chain systems which 
order magnetically due to significant interchain
couplings. These interchain interactions are the likely cause of the 
phase transition in LiVGe$_2$O$_6$ and the parallel alignment of 
neighboring spin chains suggests that they are
ferromagnetic in nature.  The strength of the interchain coupling is
estimated, using the Ginzberg-Landau theory of Scalapino, Imry, and 
Pinkus \cite{Scalapino}, to be $J_{\perp}$ $\approx$ 1.4 K (assuming
$J_{\parallel}$=45 K, four neighboring chains, and $T_N$=24 K).  These 
interchain interactions suggest that LiVGe$_2$O$_6$ is a quasi-1D system
with $J_{\perp}/J_{\parallel}$ $\approx$ 0.03 (it is important 
to note that this this mean-field
calculation results in an underestimated value of $J_{\perp}$).  
This ratio of interchain to intrachain coupling may be slightly larger than
expected given the 1D nature of the crystal structure, but is
very similar to that observed in the quasi-1D Haldane system
CsNiCl$_3$ \cite{Buyers} and certainly seems feasibile.  
Clearly, both quantum chemistry calculations
and further experiments are needed to investigate the actual magnitude
of the interchain couplings in LiVGe$_2$O$_6$.

\begin{figure}[bt]
\centerline{
\psfig{file=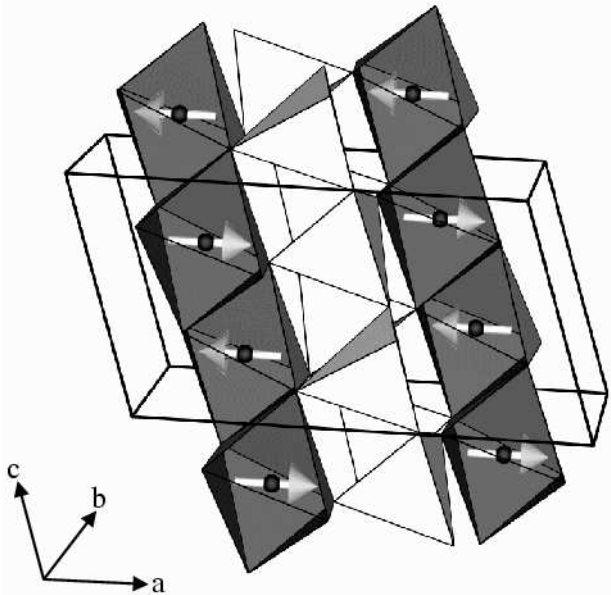,width=\columnwidth,angle=0,clip=}}
\vspace{2ex plus 1ex minus 0.5ex}
\caption{
The structure of LiVGe$_2$O$_6$ is shown for two neighboring chains
of edge-sharing VO$_6$ octahedra together with the connecting GeO$_4$
tetrahedra.  Superimposed on the VO$_6$ octahedra is the arrangement 
of spins determined from the powder neutron diffraction data showing
antiferromagnetic alignment along the chains and ferromagnetic alignment
between chains.  The spin direction could not be determined unambiguously
due to insufficient data and the indicated directions are for demonstration
purposes only.
}
\label{fig4}
\end{figure}

We cannot rule out more
exotic mechanisms for the ordering in this system and if the transition is 
discontinuous, as suggested by NMR studies, a simple ordering mediated 
by interchain coupling may be unlikely.  
However, of the theoretical ideas postulated to date, the notion of 
weakened intrachain
coupling resulting from crystal-field splitting is particularly
intriguing \cite{Gavilano,Mila}.  
Perhaps, in addition to enhancing novel interactions such
as next near-neighbor coupling or biquadratic exchange \cite{Gavilano,Mila},
this reduction of intrachain coupling also increases the relative 
strength of the interchain interactions leading
to a rather conventional ordering phase transition. In other words, 
the weakened intrachain interaction makes the system less one-dimensional. 
The mechanism 
involved in this phase transition clearly deserves further study and
several properties, such as the large magnetic gap observed
in the low temperature phase \cite{Gavilano} remain unexplained.

In summary, we have performed neutron diffraction, and complimentary x-ray
diffraction, measurements on LiVGe$_2$O$_6$ and see clear evidence for 
simple, commensurate long-range antiferromagnetic order below a 
temperature of 24 K.  In addition, we see no evidence for structural changes
at this temperature.  These results contradict the complicated picture 
of a S=1 spin-Peierls system originally proposed \cite{Millet} and seems to 
indicate surprisingly large interchain interactions.

We acknowledge fruitful discussions with D.A. Tennant, I. Affleck, and 
E. S{\o}renson.
Oak Ridge National Laboratory is managed by
UT-Battelle, LLC, for the U.S. Dept. of Energy under
contract DE-AC05-00OR22725.


%
%


\end{document}